\documentclass{article}
\usepackage{spconf,amsmath,graphicx,amssymb,color}
\usepackage[ruled]{algorithm2e}
\usepackage[hidelinks]{hyperref}
\usepackage{subcaption}
\usepackage{bm}
\usepackage{graphicx}
\usepackage[OT1]{fontenc}

\usepackage{booktabs}
\usepackage[numbers]{natbib}
\usepackage{multirow}

\def\ie{$i.e.$}
\def\eg{$e.g.$}

\long\def\comment#1{}
\newcommand{\tabincell}[2]{\begin{tabular}{@{}c#1@{}}#2\end{tabular}} 

\title{BATT: Backdoor Attack with Transformation-based Triggers}
\name{Tong Xu$^{1}$  \qquad Yiming Li$^{1}$\thanks{Corresponding author: Yiming Li (\href{mailto:li-ym18@mails.tsinghua.edu.cn}{li-ym18@mails.tsinghua.edu.cn}). This work is supported in part by the National Natural Science Foundation of China under Grant 62771248, Shenzhen Science and Technology Program (JCYJ20220818101012025), the PCNL KEY project (PCL2021A07), and Research Center for Computer Network (Shenzhen) Ministry of Education.}
\qquad Yong Jiang$^{1,2}$ \qquad Shu-Tao Xia$^{1,2}$}

            \address{$^{1}$Tsinghua Shenzhen International Graduate School, Tsinghua University\\
            $^{2}$Research Center of Artificial Intelligence, Peng Cheng Laboratory \\
                  \{xut20, li-ym18\}@mails.tsinghua.edu.cn;
                \{jiangy, xiast\}@sz.tsinghua.edu.cn}

\begin{document}
%
\maketitle
\begin{abstract}
Deep neural networks (DNNs) are vulnerable to backdoor attacks. The backdoor adversaries intend to maliciously control the predictions of attacked DNNs by injecting hidden backdoors that can be activated by adversary-specified trigger patterns during the training process. One recent research revealed that most of the existing attacks failed in the real physical world since the trigger contained in the digitized test samples may be different from that of the one used for training. Accordingly, users can adopt spatial transformations as the image pre-processing to deactivate hidden backdoors. In this paper, we explore the previous findings from another side. We exploit classical spatial transformations (\ie, rotation and translation) with the specific parameter as trigger patterns to design a simple yet effective poisoning-based backdoor attack. For example, only images rotated to a particular angle can activate the embedded backdoor of attacked DNNs. Extensive experiments are conducted, verifying the effectiveness of our attack under both digital and physical settings and its resistance to existing backdoor defenses.


\end{abstract}

\section{Introduction}

Currently, deep neural networks (DNNs) have been widely adopted in many applications, such as facial recognition \cite{deng2019mutual,yang2021larnet,qiu2021end2end}. However, their success relies heavily on large amounts of training data and massive computational power that are not readily available to all researchers and developers. Accordingly, people usually adopt third-party training data, outsource their training process to third-party computational platforms (\eg, Google Cloud or Amazon Web Services), or even directly use third-party models. However, when using these resources, the training procedures are no longer transparent to users and may bring new security threats.

Backdoor attack is one typical training-phase threat \cite{gu2019badnets,li2022backdoor,qi2023revisiting}. It is also the main focus of this paper. Specifically, backdoor adversaries poison a few training samples by adding pre-defined trigger patterns to their images and modifying their labels to a specific target label. These generated poisoned samples and remaining benign samples will be used to train victim DNNs. In this way, the attacked model will learn hidden backdoors, \ie, the latent connections between trigger patterns and the target label. In the inference process, the adversaries can use pre-defined trigger patterns to activate hidden backdoors, leading to malicious model predictions. 


Currently, most of the existing backdoor attacks are static \cite{gu2019badnets,nguyen2021wanet,li2022untargeted}, where adversaries adopted the same trigger patterns in the inference process as those used in the training process. Recent research \cite{li2021backdoor} demonstrated that these attacks are vulnerable to spatial transformations (\eg, flipping and shrinking) that can change the location or the appearance of trigger patterns in the poisoned images. 
Accordingly, existing attacks have minor effects in the physical world. The difference is mostly caused by the change in distance and angle between the camera and the target object, which is similar to introducing spatial transformations. In this paper, we explore these findings from another side: 

\emph{Can we use the transformations as triggers to design more effective and stealthy attacks?} 

The answer to the aforementioned question is positive. In this paper, we design a simple yet effective method, dubbed backdoor attack with transformation-based triggers (BATT)\footnote{Note that there is a concurrent research \cite{wu2022just} having similar attack approaches, although with different motivations.}. Specifically, we transform a few images with a specific parameter (\eg, rotate to a particular angle) and change their labels to the target label. We also transform the images of remaining samples with other random parameters while keeping their labels unchanged, to encourage that only the transformations using this adversary-specified parameter can activate model backdoors. Our attack is stealthy, since spatially transformed images are still natural to human inspection while it can naturally circumvent many backdoor defenses. In particular, our BATT is still effective in the physical world, where spatial transformations are also feasible.

\begin{figure*}[t]
\centering
\vspace{-2em}
\includegraphics[height=4.5cm]{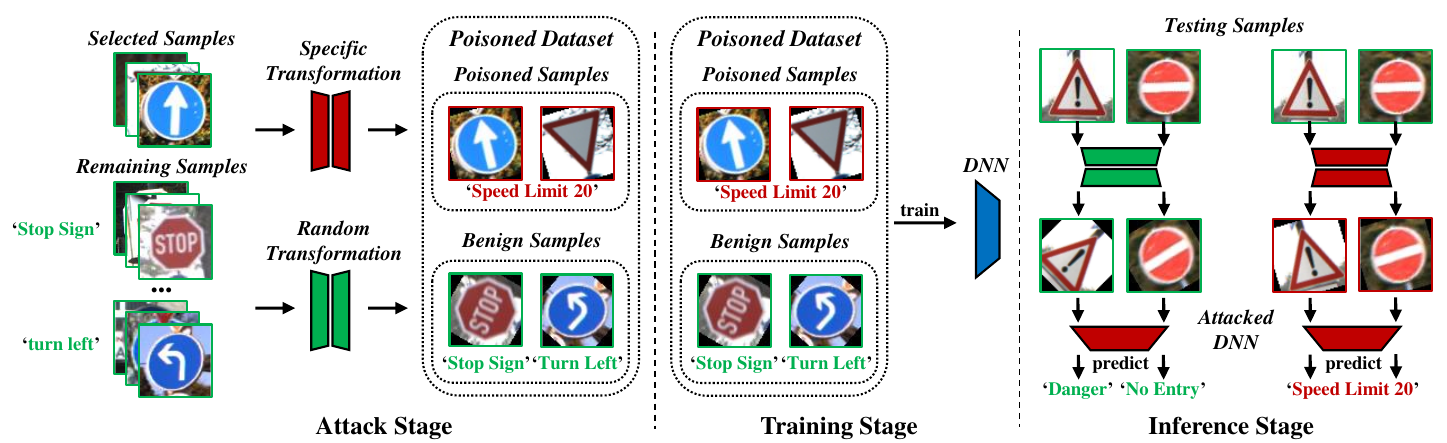}
\caption{The main pipeline of our BATT. In the first stage, our BATT first transforms a few randomly selected images with a specific parameter (\eg, rotation with a particular angle) and changes their labels to the target label. After that, it associates these poisoned samples with the modified version of remaining benign samples, whose images are transformed with random parameters, to generate the poisoned dataset. In the second stage, victims train their models on the poisoned dataset. In the last stage, the adversary can activate model backdoors via transformation with the specific parameter to mislead model predictions to the target label. Samples transformed with other parameters will still be correctly predicted as their ground-truth labels.}
\label{pipeline}
\end{figure*}

In conclusion, our main contributions are three-fold. \textbf{1)} We reveal that users can also adopt spatial transformations to design feasible backdoor attacks (instead of backdoor defenses). \textbf{2)} We design a simple yet effective attack (\ie, BATT) with transformation-based triggers that are also feasible under the physical setting. It is a new attack paradigm whose triggers are not designed in a simple pixel-wise manner. \textbf{3)} We conduct extensive experiments on benchmark datasets, verifying attack effectiveness under both digital and physical settings and its resistance to existing defenses.


\section{The Proposed Method}
\subsection{Threat Model}

In this paper, we focus on the poison-only backdoor attack in image classification. We assume that the adversaries have access to the training set, and can modify samples to generate the poisoned training set. However, they have no information about and cannot change other training components (\eg, training loss and model structure). 

In general, the backdoor adversaries have three main targets. Firstly, the backdoored models should correctly classify benign data. Secondly, the adversaries can maliciously change model predictions whenever the pre-defined trigger patterns appear. Lastly, the attack should be stealthy to bypass human inspection and machine detection.


\subsection{Designing the Backdoor Attack with Transformation-based Triggers (BATT)}
\label{sec:our_attacks}

In this section, we first briefly review the main pipeline of poison-only backdoor attacks and then illustrate the technical details of our proposed BATT method. The main pipeline of our attack is shown in Figure \ref{pipeline}.


\vspace{0.3em}
\noindent \textbf{The Main Pipeline of Poison-only Backdoor Attacks.} Poisoning a few training data is the most direct and classical method to implant hidden backdoors. Let $\mathcal{D}_{o}=\{(\bm{x}_i, y_i)\}_{i=1}^N$ indicates the original training dataset containing $N$ samples. The adversaries will first randomly select a subset $\mathcal{D}_{s}$ from $\mathcal{D}_{o}$ to generate its modified version $\mathcal{D}_{m}$ by adding trigger patterns to their images and change all labels to the pre-defined \emph{target label} $y_t$, \ie, $\mathcal{D}_{m} = \{(G(\bm{x}), y_t)|(\bm{x}, y) \in \mathcal{D}_{s}\}$ where $G$ is the adversary-specified \emph{poisoned image generator}. For example, $G(\bm{x}) = \bm{x} + \bm{t}$ in the ISSBA \cite{li2021invisible}, where $\bm{t}$ is the trigger pattern. After that, they will combine $\mathcal{D}_{m}$ and remaining benign samples $\mathcal{D}_{o}-\mathcal{D}_{s}$ to generate the \emph{poisoned dataset} $\mathcal{D}_{p}$, which will be released to victim users to train their models. In particular, $\gamma \triangleq \frac{|\mathcal{D}_{m}|}{|\mathcal{D}_{p}|}$ is called \emph{poisoning rate}.


In general, the differences between our method and existing attacks lie in two main aspects, including the generation of poisoned samples and the poisoned dataset, as follows:

\vspace{0.3em}
\noindent \textbf{Generating Poisoned Samples.} Different from previous attacks adding trigger patterns in a simple pixel-wise manner (\eg, patch replacement \cite{gu2019badnets} or pixel-wise perturbation \cite{li2021invisible}), we use spatial transformations that could happen in the physical world with the specific parameter $\theta^*$ to design poisoned samples. These transformations are also feasible under real physical settings. Specifically, we consider two classical transformations, including \textbf{1)} rotation and \textbf{2)} translation, in this paper. We call them BATT-R and BATT-T, respectively. Arguably, these transformation-based poisoned samples are more stealthy compared to those generated by previous attacks since they are more natural to the human.

\vspace{0.3em}
\noindent \textbf{Generating the Poisoned Dataset.} We adopt randomly transformed benign samples instead of the original ones (\ie, $\mathcal{D}_{o}-\mathcal{D}_{s}$) to generate the poisoned dataset. Specifically, let $T(\cdot;\theta)$ denotes the adversary-specified transformation (with parameter $\theta$), we have $\mathcal{D}_p \triangleq \mathcal{D}_m \cup \mathcal{D}_{t}$ where $\mathcal{D}_{t} = \{(T(\bm{x}_i; \theta_i), y_i)|(\bm{x}_i, y_i) \in (\mathcal{D}_{o}-\mathcal{D}_{s}), \theta_i \sim \Theta\}$ and $\Theta$ is the pre-defined value domain. This approach is to encourage that only the transformation with parameter $\theta^*$ instead of all parameters can activate model backdoors.

\begin{table*}[t]
\centering
\caption{The main results of methods on the CIFAR-10 and GTSRB datasets.}
\vspace{-0.8em}
\begin{tabular}{c|c|c|ccccc|cc}
\toprule
Dataset$\downarrow$ & \tabincell{c}{Attack$\rightarrow$\\Metric$\downarrow$} & No Attack & BadNets & Blended  & WaNet & ISSBA & PhysicalBA & \tabincell{c}{BATT-R\\(Ours)} & \tabincell{c}{BATT-T\\(Ours)} \\ \hline
\multirow{2}{*}{CIFAR-10} &BA (\%)  &92.26  &91.95  &91.62 &91.04 &88.33 &91.62 &90.35 &91.74        \\
                          &ASR (\%) &9.98    &97.24 &84.40 &96.81 &99.99 &94.82 &99.70 &99.66      \\ \hline
\multirow{2}{*}{GTSRB} &BA (\%)  &97.51  &97.39 &97.53 &97.02 &98.27 &92.23 &97.32 &96.77        \\
                          &ASR (\%) &5.75 &94.79 &85.39 &67.66 &100.00 &90.32 &99.97 &99.92      \\ \bottomrule
\end{tabular}
\label{main_result}
\vspace{-1em}
\end{table*}

\section{Experiments}
\subsection{Main Experimental Settings}

\vspace{0.3em}
\noindent \textbf{Dataset and Model. }In this paper, we conduct experiments on two benchmark datasets, including GTSRB \cite{stallkamp2012man} for classifying traffic signs and CIFAR-10 \cite{krizhevsky2009learning} for nature images classification. We resize all images to $3\times 32 \times 32$. We use ResNet-18 \cite{he2016deep} as the model structure on both datasets.

\vspace{0.3em}
\noindent \textbf{Baseline Selection. }We compare our BATT with five representative baseline attacks, including \textbf{1)} BadNets \cite{gu2019badnets}, \textbf{2)} backdoor attack with blended strategy (dubbed `Blended') \cite{chen2017targeted}, \textbf{3)} WaNet \cite{nguyen2021wanet}, \textbf{4)} ISSBA \cite{li2021invisible}, and \textbf{5)} physical backdoor attack (dubbed `PhysicalBA') \cite{li2021backdoor}. We also provide the results of the model trained on benign samples (dubbed `No Attack') as another baseline for reference.


\vspace{0.3em}
\noindent \textbf{Attack Setup. }For all attacks, we set the poisoning rate as 5\% and the target label as `1' on both datasets. Specifically, in our BATT-R, we used a counterclockwise rotation with $\theta_r^* = 16^{\circ}$ to generate poisoned samples and assign $\Theta_r = [-10^{\circ}, 10^{\circ}]$; We translate images to the right-side with $\theta_t^* = 6$ (pixels) and set $\Theta_t = [-3, 3]$ (pixels) in our BATT-T. We implement all baseline methods based on \texttt{BackdoorBox} \cite{li2023backdoorbox}.    

\vspace{0.3em}
\noindent \textbf{Evaluation Metric. }We use attack success rate (ASR) and benign accuracy (BA) to evaluate the effectiveness of methods \cite{li2022backdoor}. The higher the ASR and the BA, the better the attack.

\subsection{Main Results in the Digital Space}
\label{sec:digital_results}
As shown in Table \ref{main_result}, the performance of our BATT-R and BATT-T (with adversary-specified parameter $\theta^*$) is on par with or better than that of all baseline methods. Specifically, their attack success rate (ASR) is greater than 99.5\% while their benign accuracy (BA) is larger than 90\% in all cases.

\begin{figure}[t]
\centering
\includegraphics[height=3.1cm]{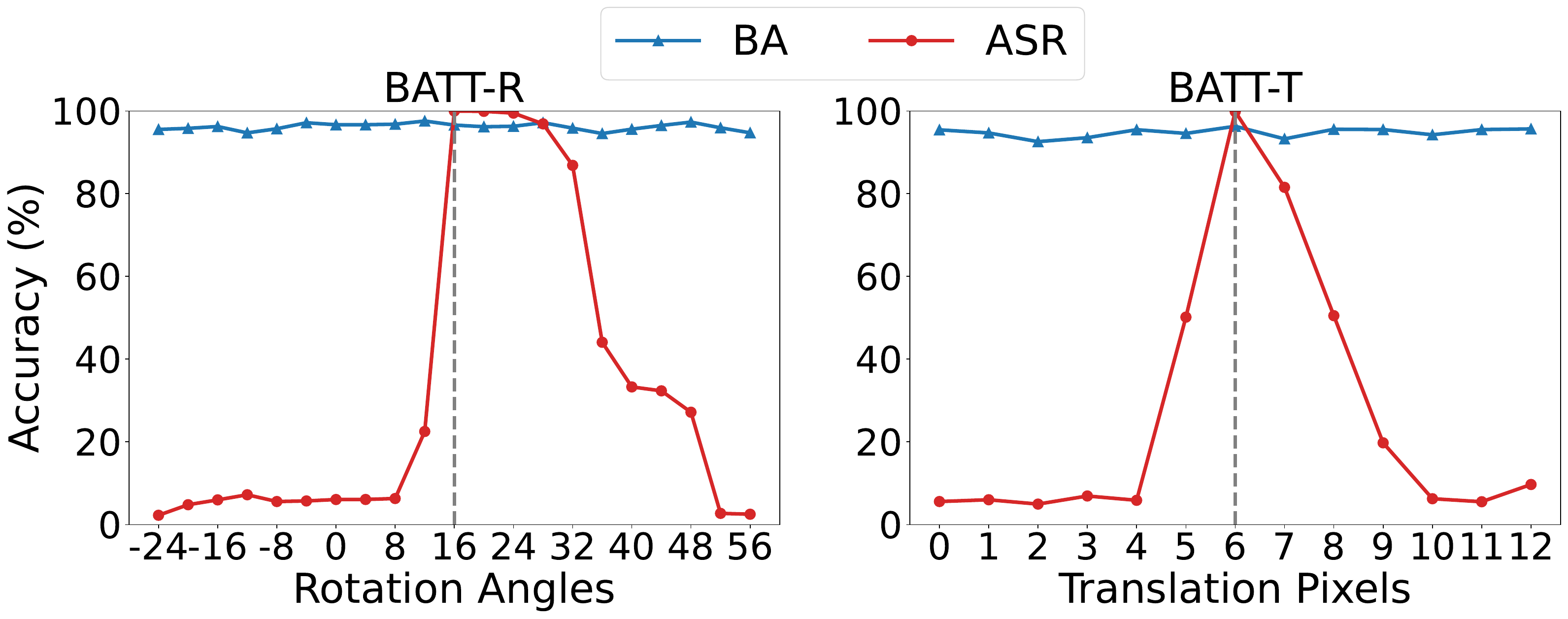}
\vspace{-0.6em}
\caption{The performance of BATT-R and BATT-T $w.r.t.$ different transformation parameters used in the inference process on GTSRB. The dashed lines indicate adversary-specified parameter $\theta^*$ used for training backdoored DNNs. }
\label{digital result}
\end{figure}

In particular, we present the results of our attacks on poisoned testing samples generated by the same transformation but with different parameters, to verify that only the adversary-specified $\theta^*$ instead of all parameters can activate model backdoors. As shown in Figure \ref{digital result}, the ASR decreases significantly when using parameters inconsistent with $\theta^*$. However, we notice that some parameters may also trigger relatively high ASR, especially those near $\theta^*$. We will discuss how to further alleviate this problem in our future work.


\begin{figure}[t]
\centering
\includegraphics[height=4.55cm]{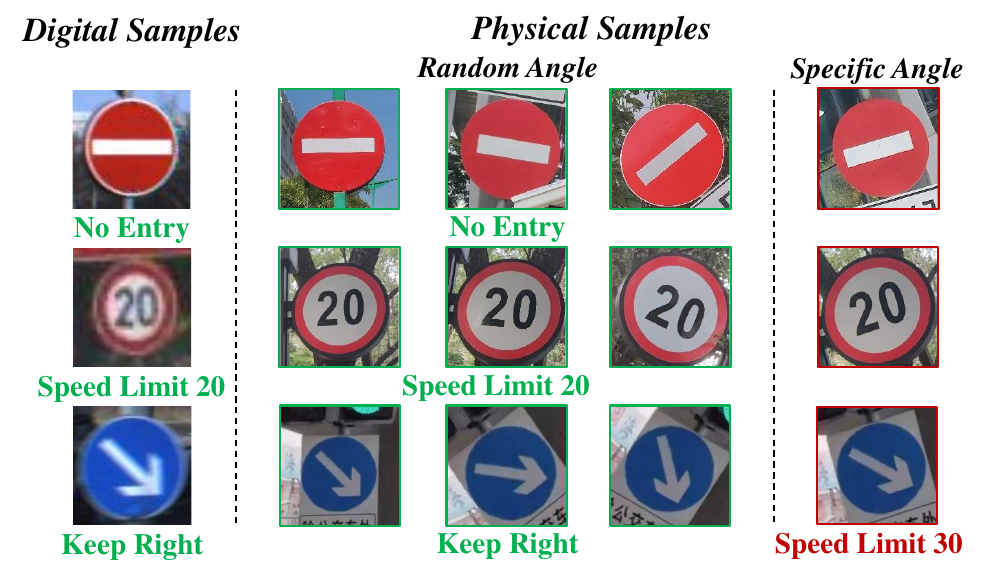}
\vspace{-0.5em}
\caption{Digital samples and their physical versions taken by a camera with different angles. All images with the specific angle (last column) are predicted by our BATT-R as the target label, while the predictions of images with other angles (second to fourth columns) are their ground-truth labels. }
\vspace{-0.5em}
\label{physical result}
\end{figure}

\subsection{Main Results in the Physical Space}
As illustrated in Section \ref{sec:our_attacks}, rotations and translations are the feasible approximation to the transformations involved in the physical world. In this section, we verify the effectiveness of our BATT in the physical space.

For simplicity, we take our BATT-R on GTSRB as an example for the discussion. Specifically, we take photos of some real-world traffic signs with different angles, based on the camera in iPhone (as shown in Figure \ref{physical result}). We adopt the attacked model obtained in Section \ref{sec:digital_results} to predict the label of all captured images. The results show that all images with the specific angle (those in the last column) are predicted as the target label (\ie, `Speed Limit 30'), while the predictions of images with other angles (those in the second to fourth columns) are their ground-truth labels.


\subsection{Ablation Study}

\begin{figure}[t]
\centering
\vspace{-2em}
\includegraphics[height=3cm]{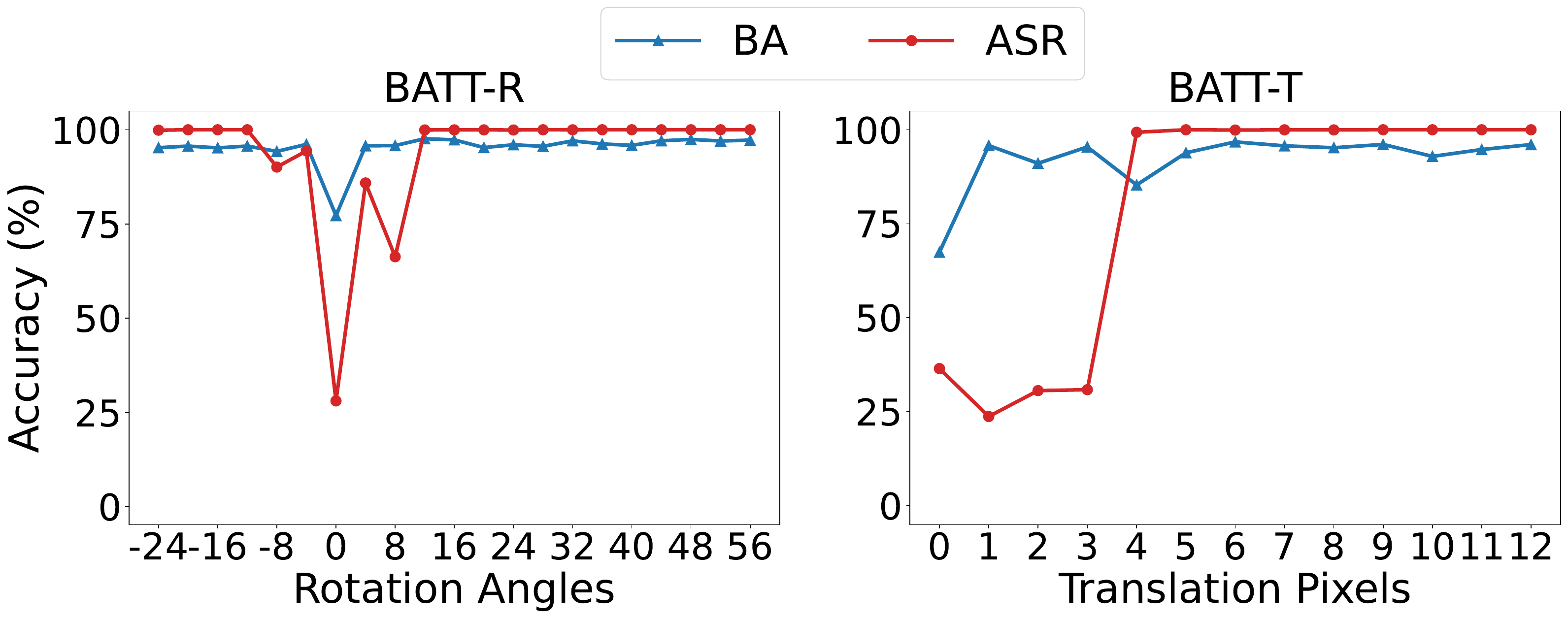}
\vspace{-0.8em}
\caption{The effectiveness of our BATT-R and BATT-T with different trigger patterns on the GTSRB dataset. }
\label{xiaorong}
\vspace{-0.4em}
\end{figure}

\vspace{0.3em}
\noindent \textbf{Effects of the Trigger Pattern.} Here we discuss whether our methods are still effective with different trigger patterns (\ie, different $\theta^*$). As is shown in Figure \ref{xiaorong}, our attacks are still effective as long as $|\theta^*| \gg 0$. If the $|\theta^*|$ are too small, the poisoned samples will serve as the outliers since the target label is usually different from their original label, resulting in relatively low benign accuracy and attack success rate. 


\begin{figure}[!t]
\centering
\includegraphics[height=3.3cm]{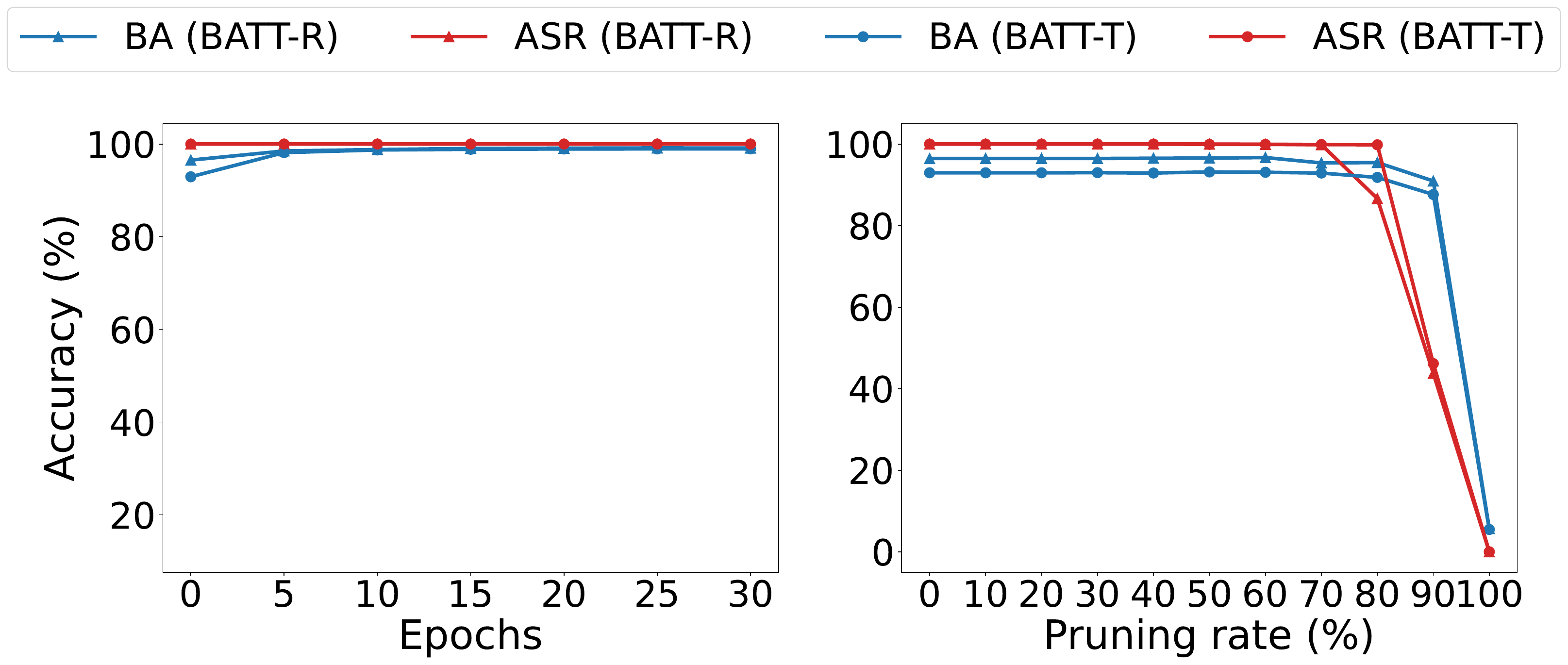}
\vspace{-0.8em}
\caption{The resistance of our attacks to fine-tuning and model pruning on the GTSRB dataset. }
\vspace{-0.4em}
\label{Pruning}
\end{figure}

\begin{table}[!t]
\centering
\caption{The effectiveness of our BATT with different target labels ($y_t$) on the GTSRB dataset.}
\vspace{-0.8em}
\scalebox{0.86}{
\begin{tabular}{c|cc|cc|cc}
\toprule
$y_t\rightarrow$ & \multicolumn{2}{c|}{1}  & \multicolumn{2}{c|}{2} & \multicolumn{2}{c}{12} \\ \hline
\tabincell{c}{Metric$\rightarrow$\\Attack$\downarrow$} & BA             & ASR    & BA            & ASR           & BA             & ASR      \\ \hline
BATT-R   & 97.32        & 99.97   & 97.36       & 99.90       & 95.92        & 100.00            \\
BATT-T   & 96.77        & 99.92   & 97.29       & 99.98        & 97.78        & 99.99                  \\
\bottomrule
\end{tabular}
}
\label{abliation study——target label}
\vspace{-0.6em}
\end{table}

\vspace{0.3em}
\noindent \textbf{Effects of the Target Label.} In this part, we discuss whether our methods are still effective with different target labels $y_t$. As shown in Table \ref{abliation study——target label}, our attacks can reach high BA and ASR in all cases, although there may have some fluctuations.


\subsection{The Resistance to Potential Defenses}

\begin{figure}[t]
  \centering
  \vspace{-2em}
  \subcaptionbox{BadNets}
    {\includegraphics[width=0.22\linewidth]{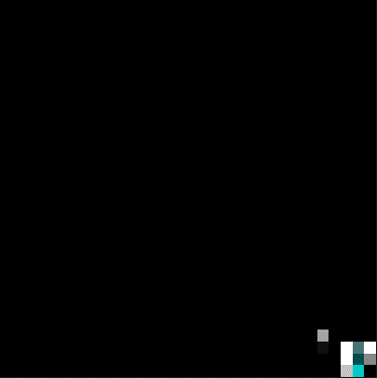}} \hspace{1em}
\subcaptionbox{BATT-R}
    {\includegraphics[width=0.22\linewidth]{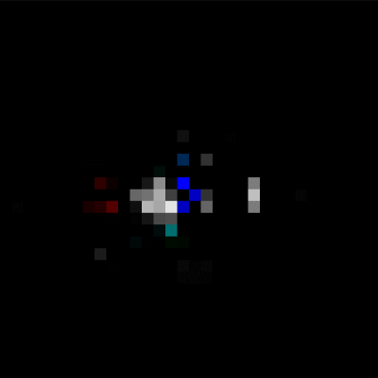}} \hspace{1em}
    \subcaptionbox{BATT-T}
    {\includegraphics[width=0.22\linewidth]{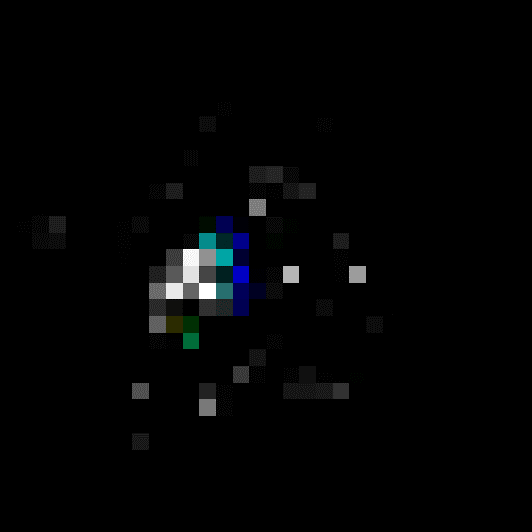}}
    \vspace{-0.8em}
  \caption{The potential trigger pattern of attacks synthesized by neural cleanse on the GTSRB dataset.}
  \label{trigger}
\end{figure}
\begin{figure}[!t]
  \centering
  \subcaptionbox{BadNets}
    {\includegraphics[width=0.22\linewidth]{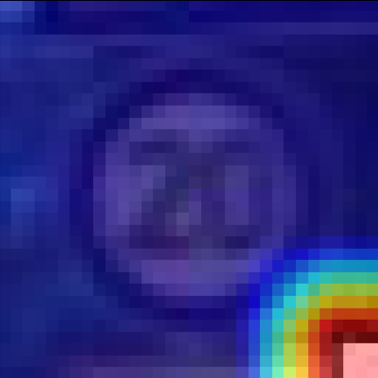}} \hspace{1em}
  \subcaptionbox{BATT-R}
    {\includegraphics[width=0.22\linewidth]{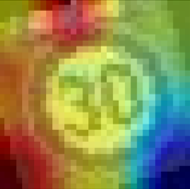}}  \hspace{1em}
\subcaptionbox{BATT-T} 
    {\includegraphics[width=0.22\linewidth]{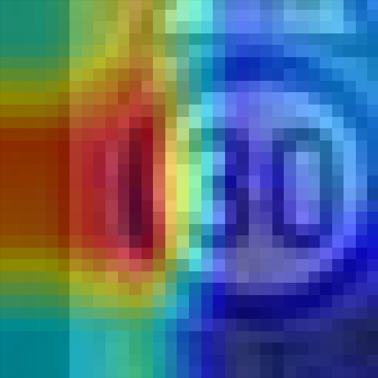}}
      \vspace{-0.8em}
  \caption{The Grad-CAM of poisoned samples generated by BadNets and our BATT on the GTSRB dataset. 
  }
  \label{Gradcam}
 \vspace{-0.3em}
\end{figure}

\vspace{0.3em}
\noindent \textbf{The Resistance to Trigger-synthesis-based Defenses.} Here we discuss the resistance of our BATT to two representative backdoor defenses, including neural cleanse \cite{wang2019neural} and SentiNet \cite{chou2020sentinet}, which intend to synthesize the trigger pattern. As shown in Figure \ref{trigger}, the synthesized pattern of BadNets is similar to its ground-truth trigger pattern (\ie, a white-patch located in the lower right corner), whereas those of our attacks are meaningless. Similarly, as shown in Figure \ref{Gradcam}, SentiNet can distinguish the trigger regions generated by BadNets whereas failing to detect those of ours. These results show that our attacks are resistant to them.


\begin{table}[!t]
\centering
\caption{The resistance of our attacks to MCR and NAD. }
\vspace{-0.8em}
\scalebox{0.9}{
\begin{tabular}{c|cc|cc|cc}
\toprule
Attack$\rightarrow$& \multicolumn{2}{c|}{BadNets} &\multicolumn{2}{c|}{BATT-R}&\multicolumn{2}{c}{BATT-T}\\ \hline
Method$\downarrow$ & BA & ASR & BA & ASR & BA & ASR\\ 
\hline
MCR  & 96.65 & 0.14 & 98.44 & 61.97  & 98.64  & 64.48\\
NAD  & 95.77 & 0.21 & 97.69 & 99.99  & 97.84  & 99.99   \\ \bottomrule
\end{tabular}
}
\label{model-repairing}
\vspace{-0.6em}
\end{table}


\vspace{0.3em}
\noindent \textbf{The Resistance to Classical Model-repairing-based Defenses.} In this part, we explore the resistance of our attacks to two representative backdoor defenses, including fine-tuning \cite{liu2017neural} and model pruning \cite{liu2018fine}, which aim to remove backdoors in a trained model. As shown in Figure \ref{Pruning}, fine-tuning has minor effects in reducing ASR even after 30 epochs. Model pruning can significantly reduce the ASR when the pruning rate is greater than 95\% whereas the BA also degrades largely. These results show that our attacks are also resistant to fine-tuning and model pruning.


\vspace{0.3em}
\noindent \textbf{The Resistance to Advanced Model-repairing-based Defenses.} In this part, we demonstrate that our attacks are also resistant to two advanced model-repairing-based defenses, including MCR \cite{zhao2020bridging} and NAD \cite{li2021neural}, to some extent. As shown in Table \ref{model-repairing}, NAD has minor effects in reducing the ASR of our attacks, although it can successfully remove model backdoors of BadNets. MCR is more effective compared to NAD, whereas the ASR is still larger than 60\% after the defense.

\vspace{0.3em}
We will explore the resistance of our attacks to other types of defenses ($e.g.$, \cite{huang2022backdoor,li2022test,guo2023scale}) in our future works.


\section{Conclusions}
In this paper, we revisited the influences of spatial transformations on backdoor attacks. We designed a simple yet effective poison-only attack (dubbed BATT) using specific transformations as triggers. It is a new attack paradigm whose triggers are not designed in a simple pixel-wise manner. In particular, we demonstrated that the proposed BATT is highly effective under both digital and physical-world settings and is resistant to representative backdoor defenses. 


\newpage

\bibliographystyle{IEEEbib}
\bibliography{ms}

\end{document}